\documentclass[12pt,reqno]{amsart}
\allowdisplaybreaks[4]
\usepackage{graphicx} 
\usepackage{amssymb}
\usepackage{color}
\usepackage{amsmath}
\usepackage{cite}
\usepackage{subfigure}
\usepackage{graphicx}
\usepackage{epstopdf}
\usepackage{bibentry}

\begin{document}
\title{On the evolution  of  a   rogue wave along the orthogonal  direction of the ($t,x$)-plane}
\author{Feng Yuan$^{1}$,  Deqin Qiu$^{1}$, Wei Liu $^2$, K. Porsezian$^3$, Jingsong He$^{1}$$^*$}
\thanks{$^*$ Corresponding author: hejingsong@nbu.edu.cn, jshe@ustc.edu.cn}
\dedicatory {$^{1}$ Department of Mathematics, Ningbo University,
Ningbo , Zhejiang 315211, P.\ R.\ China \\
$^2$ School of Mathematical Sciences, USTC, Hefei,
Anhui 230026, P.\ R.\ China\\
$^3$ Department of Physics, Pondicherry University, Puducherry 605014, India.
}


\begin{abstract}
  The localization characters of the first-order rogue wave (RW) solution $u$ of the  Kundu-Eckhaus equation is studied in this paper.
 We discover a full process  of the evolution for the contour line  with height $c^2+d$ along the orthogonal  direction of the ($t,x$)-plane for  a first-order RW $|u|^2$:   A point at height $9c^2$ generates a convex curve for $3c^2\leq d<8c^2$, whereas it  becomes a concave curve for $0<d<3c^2$, next  it reduces  to a hyperbola  on asymptotic plane (i.e. equivalently $d=0$),  and the two branches of the hyperbola become two separate convex  curves when  $-c^2<d<0$, and finally they reduce to two separate points at $d=-c^2$. Using the contour line method, the length, width, and area of the RW at height $c^2+d (0<d<8c^2)$ , i.e. above the asymptotic  plane, are defined.  We study the evolutions of three above-mentioned localization characters on $d$ through analytical and visual methods. The phase difference between the Kundu-Eckhaus and the nonlinear Schrodinger equation is also given by an explicit formula.
\end{abstract}

 \maketitle \vspace{-0.9cm}



\section{Introduction}
The nonlinear  Schr\"{o}dinger(NLS) equation, one of the most famous equation in physics, is as follows \cite{chiao, zakharov1},
\begin{equation}\label{NLSeq}
i\,u_{t}+u_{xx}+2\,|u|^2\,u=0,
\end{equation}
 which has been studied extensively from the point of view of mathematics and physics \cite{ablowitzbook,AkhmedievBook}.  Here, $u(x,t)$ is the envelope of an electric field, $t$ denotes the normalized spatial variable and $x$ is the normalized time variable. Nevertheless, in order to get
 high bit rates in optical fiber communication system, one always has to increase the intensity of the incident light field to produce ultrashort (femtosecond or even attosecond) optical pulses. In this case, a simple NLS equation is inadequate to accurately describe the propagation of light in fiber, and high-order nonlinear effects, such as third-order dispersion, self-steepening, and self-frequency shift, must be taken into consideration \cite{HMbook,agrawalbook5ed}.  To model above highly  nonlinear optical system, we have to add higher-order nonlinear terms and its derivatives into the NLS equation. It is a big challenge to do this corrections of the NLS equation without loss of the integrability.
Regarding the inclusion of above ideas, the Kundu-Eckhaus(KE) equation \cite{kundua,echaus1},  which was introduced as a cubic-quintic extension of the NLS equation and also can be reduced from several models of optics \cite{fandianyuan1,kundu1999pre,porsezian1, porsezian2,porsezian3} and
 fluid \cite{Johnson}, is one of the well-known examples.   The KE equation
is given in the form of \cite{kundua}
\begin{equation}\label{kunduechauseq}
i\,u_{t}+u_{xx}+2\,|u|^2\,u+4\,\beta^2\,|u|^4\,u-4i\,\beta\left(|u|^2\right)_{x}u=0, \quad \beta\in \mathbf{R},
\end{equation}
which contains higher nonlinearity and the Raman effect in nonlinear optics.  $\beta$ is a real constant, $\beta^2$ is the quinic nonlinear coefficient,
 and the last term  is responsible for the self-frequency shift.
The integrability aspects of  KE equation has been extensively studied by using the explicit form of the Lax pair, Painleve property \cite{clarkson}, Hamiltonian structure \cite{geng1}, soliton solutions obtained through
the Darboux transformation (DT) \cite{geng2} and  by the bilinear
method \cite{satsuma} and  by a direct method \cite{fengz1},
higher-order extension\cite{kundu2},  infinitely many conservation laws\cite{peng1}, lower-order rogue waves (RWs) given by the DT method\cite{zha1}, etc. In particular, the DT of the KE equation is not completely established \cite{geng2,zha1} because there exists an overall factor $A_A$(or $\alpha_N$) involved with complicated integrations, which  produces the difficulty in the construction of multi-fold DT.
Very recently, we overcame  this problem thoroughly by finding an explicit analytical form of the overall factor
$\prod\limits_{i=0}^{n-1}H^{[i]}$  for the n-fold DT $T_n$ (see Theorem 2.3 of Ref. \cite{qiudeqin}).  In particular, several higher-order rogue waves of the KE equation have been given explicitly in references \cite{zha1, qiudeqin, higher}.

Rogue wave is one kind of common nonlinear local waves which is also called as freak wave, monster wave, killer wave, extreme wave and abnormal wave. It is used to describe spontaneous huge ocean waves, which can lead to water walls taller than 20-30 m so that it is even a threaten to a big ship \cite{garrett,KharifPelinovskySlunyaev,Osborne1}.  The study of rogue wave has been boosted extremely by the laboratory observations in nonlinear fibers \cite{solli2,akhmediev6} and  in water tanks \cite{akhmediev7}, then rogue wave has also extended  fleetly to many fields such as plasmas, super fluids, capillary flow, Bose-Einstein condensates, the atmosphere \cite{shats,efimov,bludov,moslem,stenflo}, etc. In order to  better understand and apply the rogue wave concept in any physical system, it is necessary to  analyse the features of the  rogue wave profiles.  Especially, the squared modulus of the solution $|u|^2$ always represents a measurable quantity, optical power (or intensity).  Recently, a effective tool, contour line method, is applied to study the localization characters of the profile for rogue waves by computing the width, length and area \cite{qiudeqin, fewcycle, yongshuai,qiudeqin2}.
It is known that the contour line on the asymptotic plane is a hyperbola, while the contour line  above the asymptotic plane  is a closed curve
 and  then the width, length and area of the rogue wave can be worked out \cite{qiudeqin}. But in \cite{fewcycle, qiudeqin, yongshuai,qiudeqin2}, we  just analysed the localization characters at a given height $c^2+1$ along the orthogonal  direction of the ($t,x$)-plane, where $c^2$ is the height of asymptotic background for rogue wave $|u|^2$.  How does these localized characters are evolving  along the orthogonal  direction of the ($t,x$)-plane ?
It's known that the counter line on the height $c^2+1$ along the orthogonal  direction of the ($t,x$)-plane is a closed concave curve, for example, see Figure 7(b) of Ref. \cite{qiudeqin}.
 But whether the counter line at height $c^2+d(0\leq d \leq 8c^2$ along the orthogonal  direction of the ($t,x$)-plane, we shall explain this constraint later) is always a concave curve, and if it's not, how to find a critical height $d_c$ by changing from a concave contour line to a convex one?  Here $d$ denotes the height of counter line from the asymptotic background.
 These questions will be answered in this paper.

The rest of the paper is organized as follows. In section 2, according to the explicit expression of the first-order rogue wave solution of
the KE equation, we provide an algebraic equation to determine the counter line at height $c^2+d$ ($8c^2\geq d\geq 0$) along the orthogonal  direction of the ($t,x$)-plane, and then  find the critical height $d_c$.
The convex profile of the counter lines with height $c^2-d(c^2\geq d>0)$ along the orthogonal  direction of the ($t,x$)-plane is also discussed.  In section 3, we work out the width, length and area of the contour line lower than $d_c$, and  then  analyse the influence of
 the parameters $d$. In section 4, we consider the localization characters of the counter line higher than $d_c$.
In section 5, we give conclusions and discussions.

\section{The critical height  $d_c$}

\begin{figure}[!htbp]
\centering
\subfigure[]{\includegraphics[height=6.5cm,width=6.5cm]{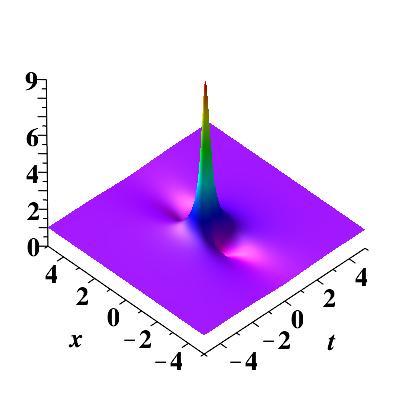}}
\qquad
\subfigure[]{\includegraphics[height=5cm,width=5cm]{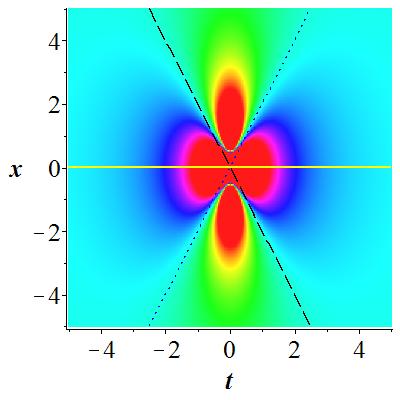}}
\caption{The profile of  the first-order RW solutions $|u_{rw}^{[1]}|^2$ with $a=0.5, c=1, \beta=0.25$. The panel (b) is the
 density plot of the panel (a).   In panel (b), $l_1$ (black, dash) and $l_2$ (blue, dot) are two  asymptotes, and  $l_3$ (yellow, solid) is
 the non-orthogonal axis  of the contour line in  equation (\ref{CL}). }\label{fig_rw2}
\end{figure}

Through the Darboux transformation, we have generated the first-order rouge wave(RW) of the KE equation \cite{qiudeqin} in the form
\begin{equation}\label{RW}
u^{[1]}_{rw}=\frac{L_{r1}}{L_{r2}}\exp\left({\rm i}(\rho+\frac{L_{r3}}{L_{r2}})\right),
\end{equation}
where
\begin{align*}
&L_{r1}=-4\,{c}^{3}{x}^{2}+ \left( -32\,\beta\,{c}^{5}+16\,a{c}^{3} \right) tx
+ \left( -64\,{\beta}^{2}{c}^{7}+64\,a\beta\,{c}^{5}-16\,{a}^{2}{c}^{3
}-16\,{c}^{5} \right) {t}^{2}+16\,it{c}^{3}+3\,c
,  \\
&L_{r2}= \left( 64\,{\beta}^{2}{c}^{6}-64\,a\beta\,{c}^{4}+16\,{a}^{2}{c}^{2}+
16\,{c}^{4} \right) {t}^{2}+ \left( 32\,\beta\,{c}^{4}-16\,a{c}^{2}
 \right) xt+4\,{c}^{2}{x}^{2}+1, \\
&L_{r3}=16\,\beta\,{c}^{2}x+ \left( 64\,{\beta}^{2}{c}^{4}-32\,a\beta\,{c}^{2}
 \right)t, \qquad
\rho=ax+(-a^{2}+4\beta^{2}c^{4}+2c^{2})t,
\end{align*}
and $a, c, \beta$ are real constant. It is trivial to find that $|u^{[1]}_{rw}|^2$ goes to $c^2$ when $|x|\rightarrow \infty$  and
$|t|\rightarrow \infty$,  which implies that  the height  of the asymptotic background is $c^2$.  $|u^{[1]}_{rw}|^2$ is a doubly-localized rational
function with a large amplitude $9c^2$ at (0,0) on ($t,x$)-plane, which reflects the two typical characters of the RW, namely, localization and large amplitude. Figure \ref{fig_rw2}
is plotted for $|u^{[1]}_{rw}|^2$ in order to show visibly the above two characteristic features.

It is shown from Figure \ref{fig_rw2} that the counter line above the
asymptotic background is reducing from a concave curve to a point when the height $c^2+d$ along the orthogonal  direction of the ($t,x$)-plane of counter line is raising from $c^2$ to $9c^2$.
Here $d$ denotes the height of contour line from the asymptotic background.
This
observation of $|u^{[1]}_{rw}|^2$ shows that  $d$ is a key parameter to control the profile of the contour line, and  strongly responsible for the existence of a  critical height $d_c$ at which we observe interesting transition from a concave contour line  to convex one.  However, there are already three parameters $a$, $c$ and $\beta$ in the RW $u^{[1]}_{rw} $, and then it is a challenging problem to illustrate analytically the role of $d$ in control of the profile.
At present the contour line method is really a useful tool to analyse the localization characters  at a given height $c^2+1$ (i.e. $d=1$) of the first-order RW solution \cite{qiudeqin, fewcycle, yongshuai,qiudeqin2}. In the following context we shall use contour line of
$|u^{[1]}_{rw}|^2$ with given values of parameters ($a,c, \beta$) to study the evolution of profile with a varying $d$.   By this method, a contour line of $|u^{[1]}_{rw}|^2$ at height $c^2+d$ ($d>0$) along the orthogonal  direction of the ($t,x$)-plane is expressed by
\begin{equation}\label{CL}
-16c^4dx^4+(-256\beta c^6d+128ac^4d)tx^3+R_1x^2+R_2x+R_3t^4+R_4t^2+8c^2-d=0,
\end{equation}
where
\begin{align*}
R_1=&\left( -1536\,{\beta}^{2}{c}^{8}d+1536\,a\beta\,{c}^{6}d-384\,{a}^{2}
{c}^{4}d-128\,{c}^{6}d \right) {t}^{2}-32\,{c}^{4}-8\,{c}^{2}d
, \\
R_2=&-4096\,{\beta}^{3}{c}^{10}d+6144\,a{\beta}^{2}{c}^{8}d-3072\,{a}^{2}
\beta\,{c}^{6}d-1024\,\beta\,{c}^{8}d+512\,{a}^{3}{c}^{4}d+512\,a{c}^{
6}d
, \\
R_3=&-4096\,{\beta}^{4}{c}^{12}d+8192\,a{\beta}^{3}{c}^{10}d-6144\,{a}^{2}{
\beta}^{2}{c}^{8}d-2048\,{\beta}^{2}{c}^{10}d\\
&+2048\,{a}^{3}\beta\,{c}^
{6}d+2048\,a\beta\,{c}^{8}d-256\,{a}^{4}{c}^{4}d-512\,{a}^{2}{c}^{6}d-
256\,{c}^{8}d
, \\
R_4=&-512\,{\beta}^{2}{c}^{8}-128\,{\beta}^{2}{c}^{6}d+512\,a\beta\,{c}^{6}+128\,a\beta\,{c}^{4}d-128\,{a}^{2}{c}^{4}+128\,{c}^{6}-32\,{a}^{2}{c}
^{2}d-32\,{c}^{4}d
.
\end{align*}
Set $d=0$ in Eq.(\ref{CL}), the contour line \cite{qiudeqin} is a hyperbola on the asymptotic plane which  has two asymptotes
\begin{equation*}
l_1:x=2(a-2\beta c^2-c)t, \qquad l_2:x=2(a-2\beta c^2+c)t,
\end{equation*}
and two non-orthogonal axes
\begin{equation*}
major \quad axis:t=0, \qquad imaginary \quad axis(l_3):x=(2a-4\beta c^2)t.
\end{equation*}
These three lines are plotted in Figure \ref{fig_rw2}(b).  As the maximum amplitude of $|u_{rw}^{[1]}|^2$ is $9c^2$, so the height of contour
line above the background must be in the interval $(c^2, 9c^2]$ or equivalently $0<d\leq 8c^2$.

Note that Eq.(\ref{CL}) is an implicit form  of the contour line. Actually, it can be  expressed explicitly  by two branches
\begin{eqnarray} \label{}
l_4:& x=-4\beta c^2t+2at+\frac{F_2}{2dc}, \label{L4} \\
l_5:& x=-4\beta c^2t+2at-\frac{F_2}{2dc}, \label{L5}
\end{eqnarray}
in which $F_2=\sqrt{-(16c^4d^2t^2+4c^2d+d^2)+4cd\sqrt{16c^4dt^2+c^2+d}}$ ($t\in [-\frac{\sqrt{d(8c^2-d)}}{4dc^2}, \frac{\sqrt{d(8c^2-d)}}{4dc^2}])$. Set $F_2=\sqrt{-F_{2A}+F_{2B}}$ and $y=t^2$, then  $F_{2B}^2-F_{2A}^2=$
$- \left( 16\,{c}^{4}y+1 \right)  \left( 16\,{c}^{4}dy-8\,{c}^{2}+d \right) \geq 0$ if  $t\in [-\frac{\sqrt{d(8c^2-d)}}{4dc^2}, \frac{\sqrt{d(8c^2-d)}}{4dc^2}])$, then $F_{2B}-F_{2A}>0$. Thus $F_2$ is a real function of $t$.  The derivatives with respect to $t$ of two branches are
\begin{equation} \label{xleq1}
l_{4}: -4\beta c^2+2a+\frac{F_1}{F_2}-x'(t)=0,\quad and \quad l_{5}: -4\beta c^2+2a-\frac{F_1}{F_2}-x'(t)=0,
\end{equation}
with $ F_1=-8c^3dt+\frac{16dc^4t}{\sqrt{16c^4dt^2+c^2+d}}$, and $x'(t)$ means $\frac{dx}{dt}$.

\begin{figure}[!htbp]
\centering
\subfigure{\includegraphics[height=12cm,width=12cm]{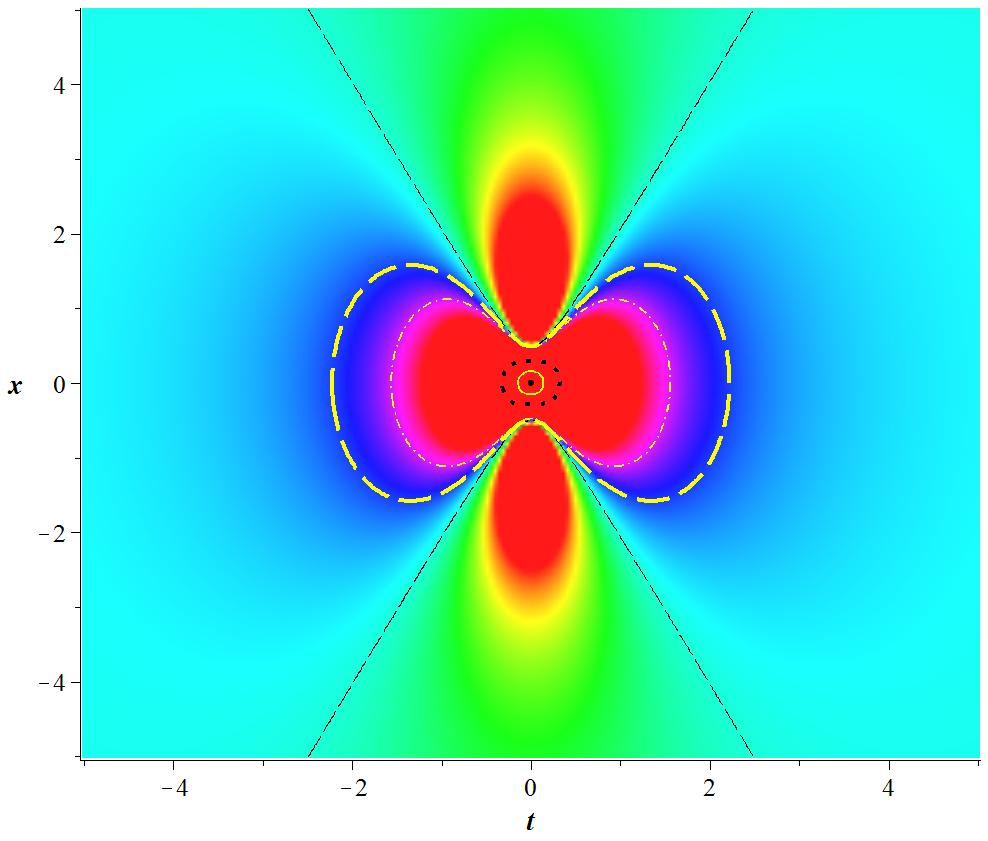}}
\caption{ The density plot and contour lines of the first-order RW solution $|u_{rw}^{[1]}|^2$ with $a=1/2, c=1, \beta=1/4$. Different values of $d$ from outside to inside: $d=0$ (longdash,thin), $d=0.1$ (dash,thick), $d=0.2$ (dashdot), $d=3$ (dot), $d=6$ (solid), $d=8$ (a point). Note that $|u_{rw}^{[1]}|^2$ reaches maximum value $9$ at (0,0), which is corresponding to the point given by contour line at $d=8$.  The parameter $d=0$ produces two branches of a hyperbola on the asymptotic plane. }\label{mdlk}
\end{figure}

We are now in a position to find $d_c$.  It can be realized by finding how many points on $l_4$ (or $l_5$) whose derivative in Eq.(\ref{xleq1})
is equal to the slope of $l_3$.  The slope of the line $l_3$ is $k=2a-4\beta c^2$. So set $x'(t)=k$,   Eq.(\ref{xleq1}) leads to two equivalent equations of $t$
\begin{equation}\label{xleq3}
-8c^3dt(\sqrt{16c^4dt^2+c^2+d}-2c)=0 \,(by \, l_{4}), 8c^3dt(\sqrt{16c^4dt^2+c^2+d}-2c)=0 \, (by \, l_{5}).
\end{equation}
By solving Eq.\eqref{xleq3} we know that $t$ has three values $0$, $\frac{\sqrt{d(3c^2-d)}}{4dc^2}$, and $-\frac{\sqrt{d(3c^2-d)}}{4dc^2}$, if $0<d<3c^2$; or it just has one value $0$ if $d \geq 3c^2$.  The former produces a concave contour line at height $c^2+d$ along the orthogonal  direction of the ($t,x$)-plane, but the latter gives  a convex one. Thus  the critical value of height is $d_c=3c^2$. In Figure \ref{mdlk}, $d_c=3$, thus contour lines associated with
 $d=0.1$ and $0.2$ (outer) are concave, but  $d=3$ and $6$ (inner) gives two convex contour lines.

\begin{figure}[!htbp]
\centering
\subfigure[]{\includegraphics[height=5cm,width=5cm]{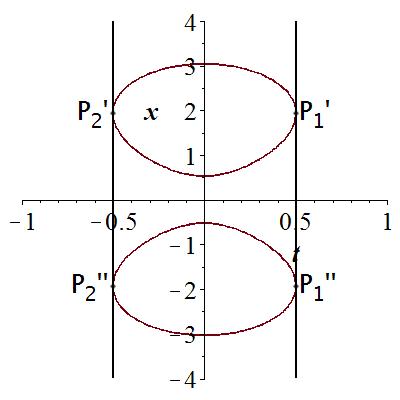}}
\caption{End points of contour lines for $|u_{rw}^{[1]}|^2$ below the asymptotic plane with height $c^2-d$.
The governing equation  of these contour lines is  Eq.(\ref{CL}) with $-d$ instead of $d$. Parameters are given by
 $a=1/2$, $c=1$, $d=1/5$, $\beta=1/4$. Four end points are $P_{1}'=(\frac{1}{2}, \frac{\sqrt{15}}{2}), P_{2}'=(-\frac{1}{2}, \frac{\sqrt{15}}{2}), P_{1}''=(\frac{1}{2}, -\frac{\sqrt{15}}{2}), P_{2}''=(-\frac{1}{2}, -\frac{\sqrt{15}}{2})$ on the ($t,x$)-plane. }\label{xia}
\end{figure}

\begin{figure}[!htbp]
\centering
\subfigure{\includegraphics[height=12cm,width=12cm]{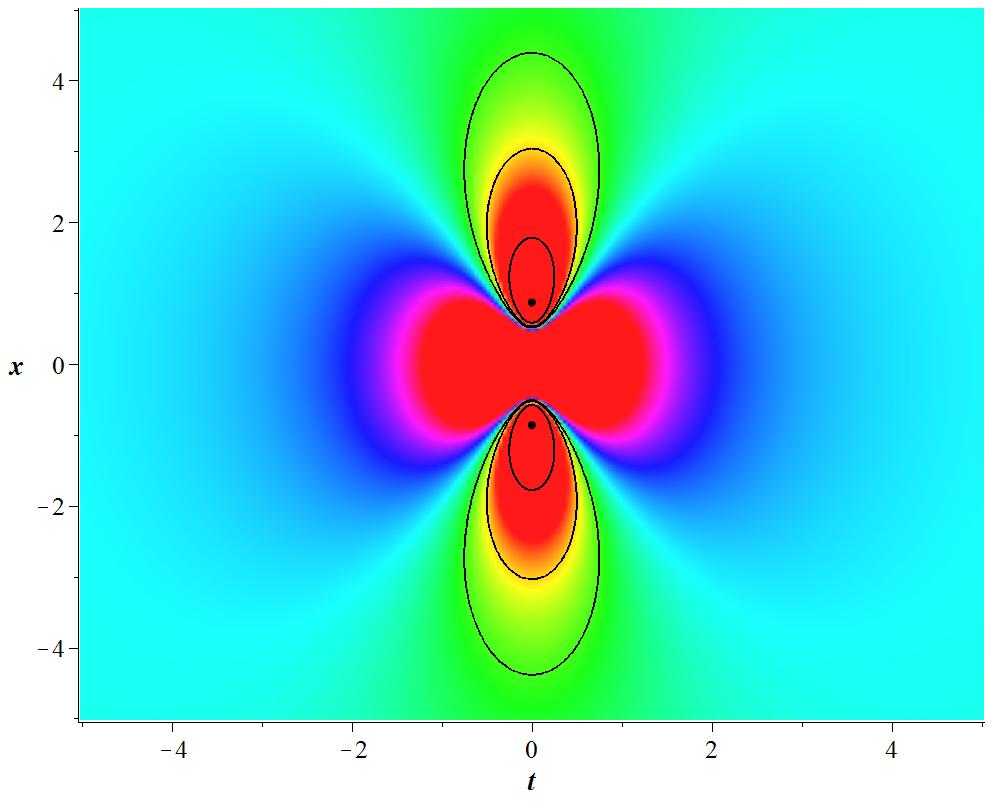}}
\caption{ The contour line of the first-order RW solution $|u_{rw}^{[1]}|^2$ with height $c^2-d$.
The governing equation  of the contour line is  Eq.(\ref{CL}) with $-d$ instead of $d$. Parameters: $a=1/2, c=1, \beta=1/4$. Different values of $d$ from outside to inside: $d=0.1$, $d=0.2$, $d=0.5$, $d=1$. Two points are given by $d=1$, which are located at
 ($0,\,{\frac {\sqrt {3}}{2}}$) and ($0,\,-{\frac {\sqrt {3}}{2}}$)  on
($t,x$)-plane. }\label{mdbelow}
\end{figure}

 For the contour line of  $|u_{rw}^{[1]}|^2$  below the asymptotic plane, all detailed calculations of this case are given in appendix.
 Its height is $c^2-d (c^2\geq d >0)$  along the orthogonal  direction of the ($t,x$)-plane and then the governing equation is Eq.(\ref{CL}) with  $-d$ instead of $d$.  This implicit equation can be expressed explicitly  by four branches, which determines a close contour line in upper
half-plane with two end points (see $P_1^{'},P_{2}^{'}$ in Figure \ref{xia}) and another  in lower half-plane with two end points (see $P_1^{''},P_{2}^{''}$ in Figure \ref{xia}). The four points are expressd by
\tiny{
\begin{eqnarray*}
&P_2^{'}=(-\frac{\sqrt{d(c^2-d)}}{4c^2d}, -\frac{(a-2\beta c^2)\sqrt{d(c^2-d)}-c^2\sqrt{3d}}{2c^2d}),
& P_1^{'}=(\frac{\sqrt{d(c^2-d)}}{4c^2d}, \frac{(a-2\beta c^2)\sqrt{d(c^2-d)}+c^2\sqrt{3d}}{2c^2d}),\\
&P_2^{''}=(-\frac{\sqrt{d(c^2-d)}}{4c^2d}, -\frac{(a-2\beta c^2)\sqrt{d(c^2-d)}+c^2\sqrt{3d}}{2c^2d}),
& P_1^{''}=(\frac{\sqrt{d(c^2-d)}}{4c^2d}, \frac{(a-2\beta c^2)\sqrt{d(c^2-d)}-c^2\sqrt{3d}}{2c^2d}).
\end{eqnarray*}
}
\normalsize
\noindent The $P_2^{'}P_1^{'}$ is parallel to   $P_2^{''}P_1^{''}$,  and their slope is the same as $l_3$. By a similar way as discussed in the last paragraph, it is not difficult to show that two separate counter lines (see Figure \ref{xia}) are convex  for all values of $d(c^2>d>0)$, unlike the counter line with  height  $c^2+d(8c^2>d>0)$ which has a critical value $d_c$ at which change occurs from a concave profile  to a convex one.
Figure \ref{mdbelow}
is plotted for the contour lines with different height $d$ below the asymptotic plane, and  the resulting curves are  convex and separate,  which gives a visual confirmation of our above results.  Two centers of valleys are $(0,\,{\frac {\sqrt {3}}{2c}})$  and $(0,\, -{\frac {\sqrt {3}}{2c}})$ (see two points in this figure), and their values are zero.

 According to the above study with the help of analytical way, we obtain a full  process  of the evolution for the counter line  with height $c^2+d$  along the orthogonal  direction of the ($t,x$)-plane for  a first-order RW  $|u_{rw}^{[1]}|^2$:  A point at height $9c^2$ generates a convex curve for $3c^2\leq d<8c^2$, whereas it  becomes a concave curve for $0<d<3c^2$, next  it reduces  to a hyperbola  on asymptotic plane (i.e. equivalently $d=0$),  and the two branches of the hyperbola become two separate convex  curves when  $-c^2<d<0$, and finally they reduce to two separate points at $d=-c^2$. Note again on ($t,x$)-plane
  that the maximum peak is located at ($0,0$), and two minimum points are located at  $(0,\,{\frac {\sqrt {3}}{2c}})$  and $(0,\, -{\frac {\sqrt {3}}{2c}})$,

\section{The contour line below the critical value}

\begin{figure}[!htbp]
\centering
\subfigure{\includegraphics[height=5cm,width=5cm]{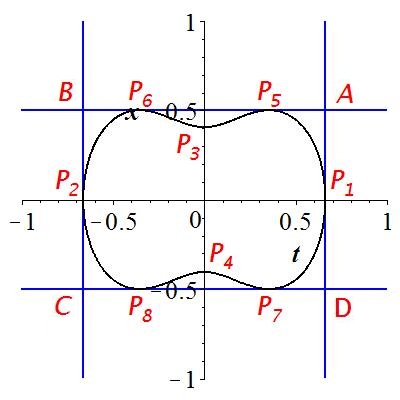}}
\caption{The tangent lines of the contour for the first-order RW solution $|u_{rw}^{[1]}|^2$ with $a=\frac{1}{2}$, $c=1$, $d=1$, $\beta=\frac{1}{4}$. }\label{jyx03}
\end{figure}

In this section, we consider the localization characters of the first-order RW when $0<d<3c^2$. According to
explicit formulas Eqs.(\ref{L4}, \ref{L5}) of two branches for contour line (see Figure \ref{jyx03}),  two end points on the ($t, x$)-plane are  $P_1=\left(\frac{\sqrt{d(8c^2-d)}}{4dc^2}, \frac{(2a-4\beta c^2)\sqrt{d(8c^2-d)}}{4dc^2}\right) $, $P_2=\left(-\frac{\sqrt{d(8c^2-d)}}{4dc^2}, -\frac{(2a-4\beta c^2)\sqrt{d(8c^2-d)}}{4dc^2}\right) $,  for all values of $a$, $c$ and $\beta$. And the equations of $BC$ and $AD$ (see Figure \ref{jyx03}) are given below,
\begin{equation}
BC: \qquad t=-\frac{\sqrt{d(8c^2-d)}}{4dc^2},  \qquad AD: \qquad t=\frac{\sqrt{d(8c^2-d)}}{4dc^2}.
\end{equation}
The length of the first-order RW solution $|u_{rw}^{[1]}|^2$ at height $d$ is defined by the distance of $P_1$ and $P_2$ \cite{fewcycle},
\begin{equation}\label{dl1}
d_{LKE}=\frac{1}{2}\sqrt{\frac{(4(a-2\beta c^2)^2+1)(8c^2-d)}{c^4d}}.
\end{equation}

For contour line in  Eq.\eqref{CL} (or equivalently  in Eqs.(\ref{L4}, \ref{L5})) with the condition $d\leq d_c$, there are  three values of $t$ at $0,\frac{\sqrt{d(3c^2-d)}}{4dc^2}, -\frac{\sqrt{d(3c^2-d)}}{4dc^2}$,  which imply six extreme points (see Figure \ref{jyx03}) at
\begin{align*}
&P_3=\left(0, \frac{\sqrt{-4c^2d-d^2+4d\sqrt{c^4+c^2d}}}{2dc}\right), \\
&P_4=\left(0, -\frac{\sqrt{-4c^2d-d^2+4d\sqrt{c^4+c^2d}}}{2dc}\right),\\
&P_5=\left({\frac {\sqrt {d \left( 3\,{c}^{2}-d \right) }}{4{c}^{2}d}}, \frac{(a-2\beta c^2)\sqrt{d(3c^2-d)}+c^2\sqrt{d}}{2c^2d}\right), \\
&P_6=\left(-{\frac {\sqrt {d \left( 3\,{c}^{2}-d \right) }}{4{c}^{2}d}}, \frac{-(a-2\beta c^2)\sqrt{d(3c^2-d)}+c^2\sqrt{d}}{2c^2d}\right), \\
&P_7=\left({\frac {\sqrt {d \left( 3\,{c}^{2}-d \right) }}{4{c}^{2}d}}, \frac{(a-2\beta c^2)\sqrt{d(3c^2-d)}-c^2\sqrt{d}}{2c^2d}\right), \\
&P_8=\left(-{\frac {\sqrt {d \left( 3\,{c}^{2}-d \right) }}{4{c}^{2}d}}, \frac{-(a-2\beta c^2)\sqrt{d(3c^2-d)}-c^2\sqrt{d}}{2c^2d}\right).
\end{align*}
By a simple calculation,  it shows $-4c^2d-d^2+4d\sqrt{c^4+c^2d}>0$ if $d\in (0, 8c^2)$, so all coordinates of $P_i(i=3,\cdots,8)$ are well defined.
Note here $d<d_c=3c^2$. The equations of $AB$ and $CD$ are as follows:
\begin{equation}\label{AB1}
AB: \qquad x-(2a-4\beta c^2)t-\frac{1}{2\sqrt{d}}=0,
\end{equation}
\begin{equation}\label{CD1}
CD: \qquad x-(2a-4\beta c^2)t+\frac{1}{2\sqrt{d}}=0.
\end{equation}
The width of the first-order RW solution  $|u_{rw}^{[1]}|^2$ at height $d$  there upon is defined by the distance of $AB$ and $CD$, which is  in the form of
\begin{equation}\label{dw1}
d_{WKE}=\frac{1}{\sqrt{d(4(a-2\beta c^2)^2+1)}}.
\end{equation}
Further the area of the first-order RW solution  $|u_{rw}^{[1]}|^2$  at height $d$ is defined by the area of the outer tangent parallelogram of the contour line, i.e.
\begin{equation}\label{s1}
S_{ABCD}=d_{LKE}d_{WKE}=\frac{\sqrt{8c^2-d}}{2c^2d}.
\end{equation}

\begin{figure}[!htbp]
\centering
\subfigure[]{\includegraphics[height=5cm,width=5cm]{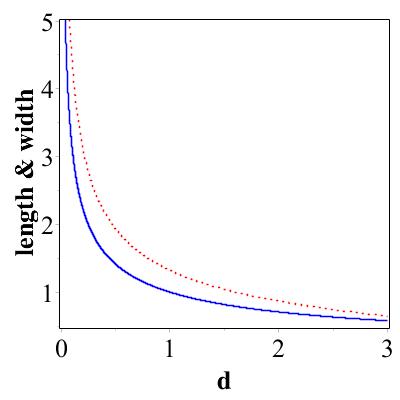}}
\subfigure[]{\includegraphics[height=5cm,width=5cm]{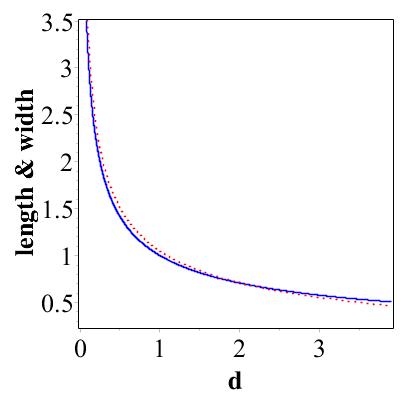}}
\subfigure[]{\includegraphics[height=5cm,width=5cm]{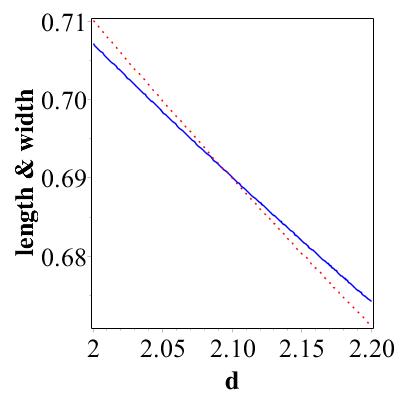}}

\caption{The minimum of length $d_{LKE_{min}}$  (red,dot) and the maximum of width $d_{WKE_{max}}$(blue,solid) of the contour line above the critical height $d_c$.   In  panel (a), $c=1$, and there is no crossing point. In panels (b, c), $c=1.3$, the latter  is the local picture of the former,  and the crossing point is given at $d=2.096$. }\label{bloweextrme}
\end{figure}

\begin{figure}[!htbp]
\centering
\subfigure[]{\includegraphics[height=5cm,width=5cm]{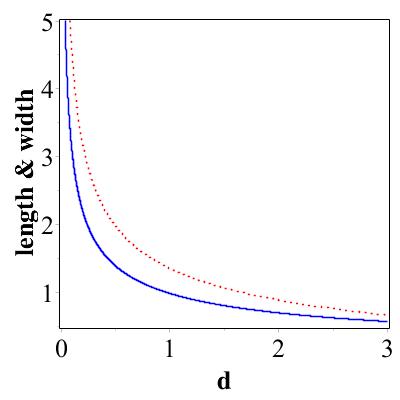}}
\subfigure[]{\includegraphics[height=5cm,width=5cm]{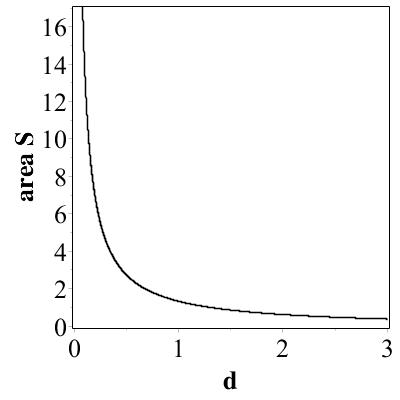}}

\caption{The length $d_{LKE}$  (red,dot) and width $d_{WKE}$(blue,solid) and the area $S$ of the contour line below the critical height $d_c$ with
parameters $ a=1/2,c=1,\beta=0.30$.   There is no crossing point in panel (a). }\label{blowlwareaond}
\end{figure}

Comparing Eq.(\ref{dl1}) with Eq.(\ref{dw1}),  there exists a condition $\beta=\frac{a}{2c^2}$ which implies a
minimum $d_{LKE_{min}}=\frac{1}{2}\sqrt{\frac{8c^2-d}{c^4d}}$ of the length  and a maximum $d_{WKE_{max}}=\frac{1}{\sqrt{d}}$ of the width.
It is interesting to note that the  condition for extreme value is independent of $d$.  If $d=4c^2(2-c^2)$ and $c\in (\frac{\sqrt{5}}{2}, \sqrt{2})$,  $d_{LKE_{min}}=d_{WKE_{max}}$. In Figure \ref{bloweextrme}(a), there is no crossing point because $c\notin (\frac{\sqrt{5}}{2}, \sqrt{2})$, but Figure \ref{bloweextrme}(b) has one crossing point at $2.096$.
\noindent For  $\beta\not=\frac{a}{2c^2}$,  we do not analyse  the condition of $d_{LKE}=d_{WKE}$ because of its complexity, and then Figure \ref{blowlwareaond} is plotted to show the evolution of  three localization characters on $d$.

\section{The contour line above the critical value}
In this section, we consider the localization characters of the first RW when $3c^2\leq d< 8c^2$ along the orthogonal  direction of the ($t,x$)-plane. Within this region, the contour line also has two end points $Q_1=\left(\frac{\sqrt{d(8c^2-d)}}{4dc^2}, \frac{(2a-4\beta c^2)\sqrt{d(8c^2-d)}}{4dc^2}\right) $, $Q_2=\left(-\frac{\sqrt{d(8c^2-d)}}{4dc^2}, -\frac{(2a-4\beta c^2)\sqrt{d(8c^2-d)}}{4dc^2}\right) $, on the ($t, x$) plane of all values of $a$ , $c$ and $\beta$. And the equations of $B^{'}C^{'}$ and $A'D'$ is given below,
\begin{equation}
B^{'}C^{'}: \qquad t=-\frac{\sqrt{d(8c^2-d)}}{4dc^2},  \qquad A^{'}D^{'}: \qquad t=\frac{\sqrt{d(8c^2-d)}}{4dc^2}.
\end{equation}

\begin{figure}[!htbp]
\centering
\subfigure{\includegraphics[height=5cm,width=5cm]{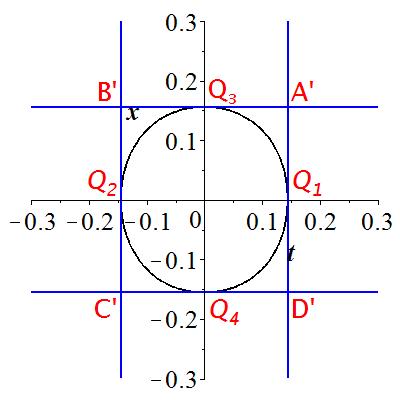}}
\caption{The tangent lines of the contour of the first-order RW solution $|u_{rw}^{[1]}|^2$ with  $a=0.5$, $c=1$, $d=6$, $\beta=0.25$ }\label{jyx4}
\end{figure}

The length of the first-order RW solution $|u_{rw}^{[1]}|^2$  is the distance of $Q_1$ and $Q_2$ (see Figure\ref{jyx4} ), i.e.
\begin{equation}\label{lke}
d_{LKE}^{'}=\frac{1}{2}\sqrt{\frac{(4(a-2\beta c^2)^2+1)(8c^2-d)}{c^4d}}.
\end{equation}
The length in Eq.(\ref{lke})  is the same as  the counter line below the critical value, but the other two localization  characters (i.e. width and area) are  different for two cases. If $d\geq 3c^2$ , the Eq.\eqref{xleq3} has only one solution, i.e. $t=0$. So points $P_5$ and $P_6$ are
approaching  to  point $P_3=Q_3=\left(0,\frac{\sqrt{-4c^2d-d^2+4d\sqrt{c^4+c^2d}}}{2dc}\right)$ when $d$ is passing by  $d_c$  from
a lower height. Similarly,  points  $P_7$ and $P_8$ are approaching  to  point $P_4=Q_4=\left(0,-\frac{\sqrt{-4c^2d-d^2+4d\sqrt{c^4+c^2d}}}{2dc}\right)$.  We can work out the equations of $A^{'}B^{'}$ and $C^{'}D^{'}$
(see Figure\ref{jyx4}) as
 \begin{equation}
 A^{'}B^{'}: \qquad  x-(2a-4\beta c^2)t+\frac{\sqrt{-4c^2d-d^2+4d\sqrt{c^4+c^2d}}}{2dc}=0,
 \end{equation}
\begin{equation}
 C^{'}D^{'}: \qquad  x-(2a-4\beta c^2)t-\frac{\sqrt{-4c^2d-d^2+4d\sqrt{c^4+c^2d}}}{2dc}=0.
 \end{equation}
The width of the  rogue wave $|u_{rw}^{[1]}|^2$  at height $d$ is defined by the distance of above two lines, which is
\begin{equation}\label{wke}
d_{WKE}^{'}=\sqrt{\frac{4c\sqrt{c^2+d}-4c^2-d}{c^2d(4(a-2\beta c^2)^2+1)}}.
\end{equation}
The area of the rogue wave $|u_{rw}^{[1]}|^2$  at height $d$ is defined by the area of the outer tangent parallelogram (see Figure \ref{jyx4}) of its contour line  at same height, which is given by
\begin{equation}\label{ske}
S_{ABCD}^{'}=d_{LKE}^{'}d_{WKE}^{'}=\frac{\sqrt{(4c\sqrt{c^2+d}-4c^2-d)(8c^2-d)}}{2c^3d}.
\end{equation}

\begin{figure}[!htbp]
\centering
\subfigure[]{\includegraphics[height=5cm,width=5cm]{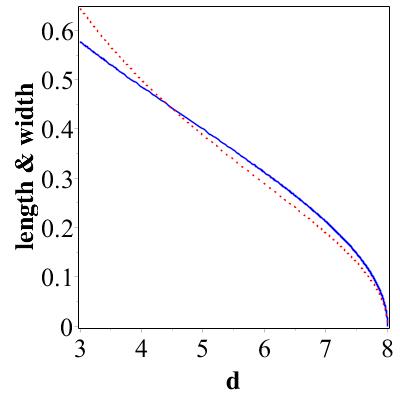}}
\subfigure[]{\includegraphics[height=5cm,width=5cm]{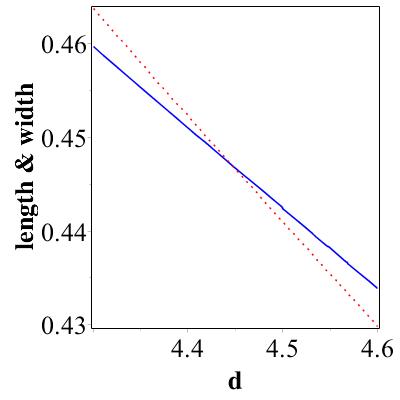}}
\subfigure[]{\includegraphics[height=5cm,width=5cm]{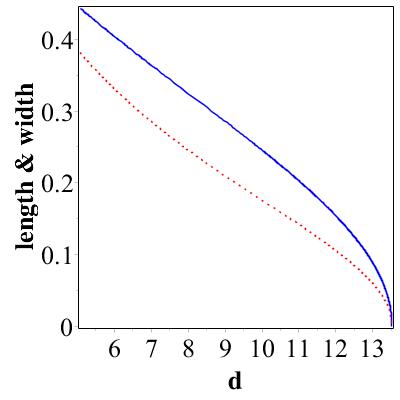}}
\caption{The minimum of length $d^{'}_{LKE_{min}}$  (red,dot) and the maximum of width $d^{'}_{WKE_{max}}$(blue,solid) of the contour line above the critical height $d_c$.  In panels (a, b), $c=1$, the latter  is the local picture of the former,  and the crossing point is given at $d=4.444$.  In
 panel (c), at $c=1.3$, there is no crossing point.}\label{aboveextrme}
\end{figure}

 Eqs. (\ref{lke},\ref{wke}) show that the length and the width reach the extreme at the same time when $\beta=\frac{a}{2c^2}$ as well.
 Specifically, the maximum $d^{'}_{WKE_{max}}=\sqrt{\frac{4c\sqrt{c^2+d}-4c^2-d}{c^2d}}$ and   $d^{'}_{LKE_{min}}=\frac{1}{2}\sqrt{\frac{8c^2-d}{c^4d}}$. A simple calculation leads to
 $d^{'}_{WKE_{max}}=d^{'}_{LKE_{min}}$ if $d=\frac{8c^2(4c^2+1)}{(4c^2-1)^2}$ and $c\in (\frac{\sqrt{3}}{2}, \frac{\sqrt{5}}{2} ]$.
 Figure \ref{aboveextrme} shows the evolution of $d^{'}_{LKE_{min}}$ and $d^{'}_{WKE_{max}}$ on $d$. There is no crossing point in Figure
 \ref{aboveextrme}(c) because $c\notin (\frac{\sqrt{3}}{2}, \frac{\sqrt{5}}{2} ] $.

\begin{figure}[!htbp]
\centering
\subfigure[]{\includegraphics[height=5cm,width=5cm]{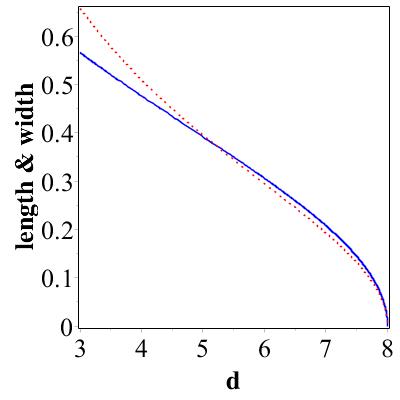}}
\subfigure[]{\includegraphics[height=5cm,width=5cm]{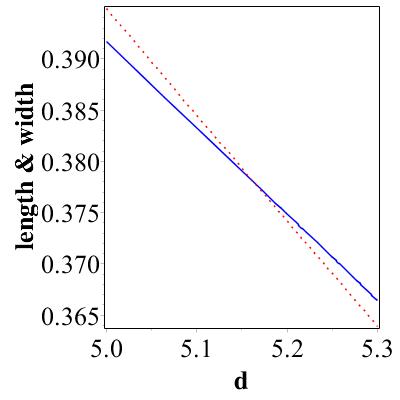}}
\subfigure[]{\includegraphics[height=5cm,width=5cm]{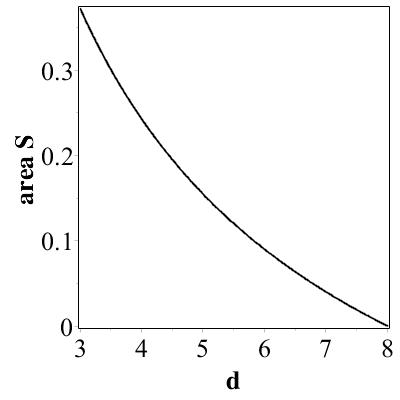}}
\caption{The length $d_{LKE}$  (red,dot) and width $d_{WKE}$(blue,solid) and the area $S$ of the contour line above the critical height $d_c$ with
parameters $ a=1/2,c=1,\beta=0.30$.   There is a crossing point at $d=5.163$ in panel (a), panel(b) is the local picture of panel(a).}\label{abovelwareaond}
\end{figure}

\begin{figure}[!htbp]
\centering
\subfigure[]{\includegraphics[height=5cm,width=5cm]{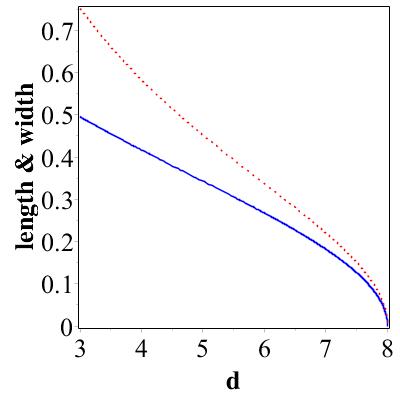}}
\caption{The length $d_{LKE}$  (red,dot) and width $d_{WKE}$(blue,solid) of the contour line above the critical height $d_c$ with
parameters $ a=1/2,c=1,\beta=0.40$.   There is no crossing point. }\label{abovelwondnointersect}
\end{figure}

\noindent For  $\beta\not=\frac{a}{2c^2}$,  we do not analyse this condition $d^{'}_{LKE}=d^{'}_{WKE}$ because of its complexity, and then Figure \ref{abovelwareaond} is plotted to merely show the evolution of  three localization characters on $d$.
In order to show the existence of $d^{'}_{LKE}\not=d^{'}_{WKE}$,  Figure \ref{abovelwondnointersect} is plotted for this case.

From our above investigations, it is clear that by suitably manipulating the value of ``$d$", one can choose suitable area, length, width and amplitude of the rogue wave type optical pulses. To the best our knowledge, so far only variable coefficients of the nonlinear evolution equations have been used to manipulate the optical rogue waves. By suitably correlating the value of $d$ in terms of the experimental parameter, our results may be useful for the experimental realization of the control of the rogue wave. {Moreover, the area $S$ and extreme values of
length and width are independent of $\beta$, which means that  higher-order terms in the KE equation do not have contribution of these three
localization characters. }

\section{Conclusions}
In this paper, we use the contour line method to study the localization characters of the first-order RW solution $|u_{rw}^{[1]}|^2$ of the KE equation. A full  evolution process for the contour line  with height $c^2+d$  along the orthogonal  direction of the ($t,x$)-plane of the  RW $|u_{rw}^{[1]}|^2$ can be summarized as follows:
  A point at height $9c^2$ generates a convex curve for $3c^2\leq d<8c^2$, whereas it  becomes a concave curve for $0<d<3c^2$, next  it reduces  to a hyperbola  on asymptotic plane (i.e. equivalently $d=0$),
  and the two branches of the hyperbola become two separate convex  curves when  $-c^2<d<0$, and finally they reduce to two separate points at $d=-c^2$.

 Analytical formulas of length, width and area of  the  RW $|u_{rw}^{[1]}|^2$ with height
$d (8c^2>d>0)$ are discussed according to two cases: $3c^2>d>0$ (below the critical value $d_c$)  and $8c^2>d \geq 3c^2$ (above the critical value $d_c$).
All of them are monotonically decreasing functions of $d$.
In the  above-mentioned three characters, only length for two cases  has a same formula, and
length and width are equal for some special values of $d$ and $c$.  Under condition $\beta=\frac{a}{2c^2}$,  the length reaches its minimum, but width gives a maximum.
  The main differences of two cases are listed by:
\begin{itemize}
\item Contour line is concave for the former, but convex for the latter. The critical value  of height for the turning between convex and concave profile is $d_c=3c^2$.
\item Different formulas for width.
\item Different formulas for area.
\item Different intervals of $c$ to get equal extreme values of length and width: $c \in (\frac{\sqrt{5}}{2},\sqrt{2})$ for former, but
$c \in (\frac{\sqrt{3}}{2}, \frac{\sqrt{5}}{2}]$ for the latter.
\end{itemize}
Our in-depth analysis on the contour line  will be useful for experimentalist to control the patterns of the rogue wave ultra-short optical light pulses.

\begin{figure}[!htbp]
\centering
{\includegraphics[height=5cm,width=5cm]{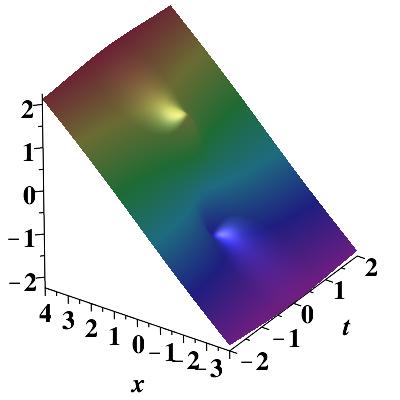}}
\caption{ The phase difference $\Delta \theta$ between rogue wave solutions of the KE and NLS with parameters $a=1, c=1$ and $\beta=0.25$.}\label{phasedifference}
\end{figure}

\begin{figure}[!htbp]
\centering
\subfigure[]{\includegraphics[height=5cm,width=5cm]{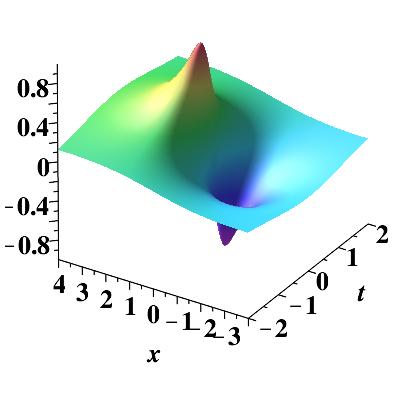}}
\subfigure[]{\includegraphics[height=5cm,width=5cm]{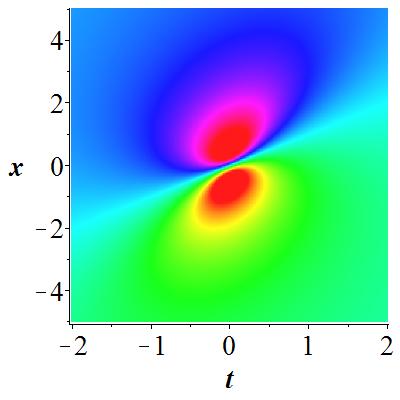}}
\subfigure[]{\includegraphics[height=5cm,width=5cm]{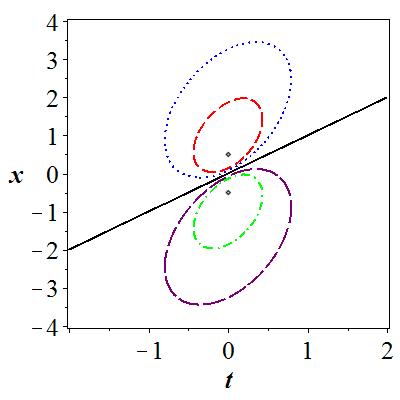}}
\caption{ The phase difference  $\widetilde{\Delta  \theta}$  with parameters $a=1, c=1$ and $\beta=0.25$. Panel (b) is the density plot of (a). Panel (c) is plotted for the contour lines at different heights of (a). In panel (c), the straight line (solid black) is plotted for $d=0$. In upper plane of (c) devided by the straight line, different values of $d$ from outside to inside: d=0.5(red dot),0.3(blue dash). In lower plane of (c), different values of $d$ from outside to inside: d=-0.3 (purple long dash), -0.5 (green dash dot).  The point in upper plane is the maximum attained at  $d=4\beta c=1$, but the point in lower plane is the minimum attained at $d=-4\beta c=-1$.   }\label{phase}
\end{figure}

Since the KE is equivalent to the NLS by a nonlinear transformation $q=u\exp(2i\beta \int|u|^2dx)$ \cite{kundua} regarding a solution $u$ of the former and a solution $q$ of the latter,
in order to show the true characteristics of the KE equation, it is also interesting and necessary  to study the phase of the first-order rogue wave $u^{[1]}_{rw}$  of the KE.
 This has been done partially  by studying the real part of the first-order RW Re$u^{[1]}_{rw}$ in Ref.\cite{qiudeqin}, which has shown that Re$u^{[1]}_{rw}$  has a central pattern around point ($0,0$)  and  alternately appeared  parallels (see Figure 3 in Ref.\cite{qiudeqin}).
 This observation implies clearly that Re$u^{[1]}_{rw}$ is nonlocal.   Thus, it is more essential to study the localized property of a phase  difference  $\Delta \theta=2\beta\int|u|^2dx$ \cite{qiudeqin} between above two solutions.
Substituting $u=u^{[1]}_{rw}$ into $\Delta \theta$, it becomes
\begin{eqnarray}
&\Delta \theta=&2\beta c^2x+\frac{32c^2\beta t(2\beta c^2-a)+16c^2\beta x}{16c^2t(x+c^2t)(2\beta c^2-a)^2+4c^2x^2+16c^4t^2+1}  \nonumber ,
\end{eqnarray}
which is plotted in Figure \ref{phasedifference}( see also in  Figure 4 of Ref.\cite{qiudeqin}).  There  exist a remarkable peak and hollow in the profile of $\Delta \theta$. If $t$ is large sufficiently,  $\Delta \theta=2\beta c^2 x$ which gives an asymptotic plane.
So, in order to illustrate clearly  the localized property of the $\Delta \theta$, it is better to study the contour lines of  $ \widetilde{\Delta  \theta}= \Delta \theta- 2\beta c^2 x$ by removing the oblique asymptotic background, namely
\begin{eqnarray}
&\widetilde{\Delta  \theta} =\frac{32c^2\beta t(2\beta c^2-a)+16c^2\beta x}{16c^2t(x+c^2t)(2\beta c^2-a)^2+4c^2x^2+16c^4t^2+1}  \nonumber.
\end{eqnarray}
$\widetilde{\Delta  \theta}$ is plotted in Figures \ref{phase}(a,b), which shows $\widetilde{\Delta  \theta}$  is doubly localized  in both $x$ and $t$.   A simple calculation gives that $\widetilde{\Delta  \theta}|_{max}=4\beta c$ at point ($t=0,x=\frac{1}{2c}$)  and  $\widetilde{\Delta  \theta}|_{min}=-4\beta c$ at point ($t=0,x=-\frac{1}{2c}$),
and the height of the  asymptotic plane for  $\widetilde{\Delta  \theta}$  is zero.  Setting $d \in [-4\beta c, 4\beta c]$, an algebraic equation of the contour line at height $d$ is given by
\begin{eqnarray}
& & 256\,{c}^{6}d \left( 4\,{\beta}^{2}{c}^{4}-4\,a\beta\,{c}^{2}+{a}^{2}+
{c}^{2} \right)t^4+256c^6(a-\beta c^2)(dx-2\beta)t^3+16\,{c}^{2} (4\,{\beta}^{2}{c}^{4}d
\nonumber   \\
& & +4\,{c}^{4}d{x}^{2}-16\,\beta
\,{c}^{4}x-4\,a\beta\,{c}^{2}d+{a}^{2}d+2\,{c}^{2}d)
t^2+16c^2(a-\beta c^2)(dx-2\beta)t+ d=0.
\end{eqnarray}
In particular, the contour line is a straight line on asymptotic plane with height zero, namely $ x=2(-2 \beta c^2+a)t$. By a similar analysis of the contour line for $u^{[1]}_{rw}$ ,  we find a full evolution of the contour line of phase difference $\widetilde{\Delta \theta}$ as follows:
  A point at height $4\beta c$ generates a convex curve for $0 < d< 4\beta c$, next  it reduces  to a straight line   on asymptotic plane (i.e. equivalently $d=0$),  and then this straight line  becomes  a convex  curve when  $-4\beta c<d<0$, and finally it
   reduced to a point at $d=-4 \beta c$, which is confirmed by Figure \ref{phase}(c). Clearly, the evolution of contour line of $\widetilde{\Delta  \theta}$  is different from the contour line of the $u^{[1]}_{rw}$.

{\bf Acknowledgments} {\noindent \small This work is supported by the NSF of China under Grant No.11271210 and K.C. Wong Magna Fund in Ningbo University.  J. He thanks sincerely Prof. A.S. Fokas for arranging the visit to Cambridge University in 2012-2015 and for many useful discussions. K.P. thanks the DST,NBHM, IFCPAR,DST-FCT and CSIR, Government of India, for the financial support through major projects. We thank referees for their helpful suggestions on the early version of this paper.}

\section*{Appendix}
In this appendix, a few additional explanations about  the contour line below the asymptotic background will be given.
For this case, the height of the contour line is  $c^2-d$ ($0<d<c^2$) along the orthogonal  direction of the ($t,x$)-plane, which is a closed curve
defined by Eq.(\ref{CL}) with $-d$ instead of $d$, i.e.

\begin{equation}\label{CLB}
16c^4dx^4+(256\beta c^6d-128ac^4d)tx^3+S_1x^2+S_2x+S_3t^4+S_4t^2+8c^2+d=0,
\end{equation}
where
\begin{align*}
S_1=&\left( 1536\,{\beta}^{2}{c}^{8}d-1536\,a\beta\,{c}^{6}d+384\,{a}^{2}
{c}^{4}d+128\,{c}^{6}d \right) {t}^{2}-32\,{c}^{4}+8\,{c}^{2}d
, \\
S_2=&4096\,{\beta}^{3}{c}^{10}d-6144\,a{\beta}^{2}{c}^{8}d+3072\,{a}^{2}
\beta\,{c}^{6}d+1024\,\beta\,{c}^{8}d-512\,{a}^{3}{c}^{4}d-512\,a{c}^{
6}d
, \\
S_3=&4096\,{\beta}^{4}{c}^{12}d-8192\,a{\beta}^{3}{c}^{10}d+6144\,{a}^{2}{
\beta}^{2}{c}^{8}d+2048\,{\beta}^{2}{c}^{10}d\\
&-2048\,{a}^{3}\beta\,{c}^
{6}d-2048\,a\beta\,{c}^{8}d+256\,{a}^{4}{c}^{4}d+512\,{a}^{2}{c}^{6}d+
256\,{c}^{8}d
, \\
S_4=&-512\,{\beta}^{2}{c}^{8}+128\,{\beta}^{2}{c}^{6}d+512\,a\beta\,{c}^{6}-128\,a\beta\,{c}^{4}d-128\,{a}^{2}{c}^{4}-128\,{c}^{6}-32\,{a}^{2}{c}
^{2}d+32\,{c}^{4}d
.
\end{align*}
Actually,  Eq. \eqref{CLB} can be  expressed explicitly  by the following four branches
 \begin{eqnarray}\label{}
l_{01}: &x=-4\beta c^2t+2at+\frac{G_1}{2dc}, \label{l01} \,\,  l_{02}: x=-4\beta c^2t+2at-\frac{G_1}{2dc},\label{l02}
\\
l_{03}: &x=-4\beta c^2t+2at+\frac{G_2}{2dc}, \label{l03} \,\,  l_{04}: x=-4\beta c^2t+2at-\frac{G_2}{2dc},\label{l04}
\end{eqnarray}
in which
\begin{align*}
&G_1=\sqrt{-16c^4d^2t^2+4c^2d-d^2+4cd\sqrt{-16c^4dt^2+c^2-d}}, \\ &G_2=\sqrt{-16c^4d^2t^2+4c^2d-d^2-4cd\sqrt{-16c^4dt^2+c^2-d}}.
\end{align*}
and $t\in [-\frac{\sqrt{d(c^2-d)}}{4dc^2}, \frac{\sqrt{d(c^2-d)}}{4dc^2}]$.
Set $G_A=-16c^4d^2t^2+4c^2d-d^2, G_B=4cd\sqrt{-16c^4dt^2+c^2-d}, G_1=\sqrt{G_{A}+G_{B}},\,G_2=\sqrt{G_{A}-G_{B}}$ and $y=t^2$, then
 $G_A>0$ and  $G_B>0$ when  $t\in (-\frac{\sqrt{d(c^2-d)}}{4dc^2}, \frac{\sqrt{d(c^2-d)}}{4dc^2})$, and
$G_{A}^2-G_{B}^2={d}^{3} \left( 16\,{c}^{4}y+1 \right)  \left( 16\,{c}^{4}dy+8\,{c}^{2}
+d \right)\geq 0$. Thus $G_A-G_B>0$ when $t\in (-\frac{\sqrt{d(c^2-d)}}{4dc^2}, \frac{\sqrt{d(c^2-d)}}{4dc^2})$, which implies
$G_1$ and $G_2$ are real functions of $t$. According to the four branches, four end points are represented explicitly by
\begin{align*}
&P_1^{'}=(\frac{\sqrt{d(c^2-d)}}{4c^2d}, \frac{(a-2\beta c^2)\sqrt{d(c^2-d)}+c^2\sqrt{3d}}{2c^2d}),\\
&P_2^{'}=(-\frac{\sqrt{d(c^2-d)}}{4c^2d}, -\frac{(a-2\beta c^2)\sqrt{d(c^2-d)}-c^2\sqrt{3d}}{2c^2d}),
\\
&P_1^{''}=(\frac{\sqrt{d(c^2-d)}}{4c^2d}, \frac{(a-2\beta c^2)\sqrt{d(c^2-d)}-c^2\sqrt{3d}}{2c^2d}),\\
&P_2^{''}=(-\frac{\sqrt{d(c^2-d)}}{4c^2d}, -\frac{(a-2\beta c^2)\sqrt{d(c^2-d)}+c^2\sqrt{3d}}{2c^2d}).
\end{align*}

Now we will prove that the contour line below the asymptotic background is always convex.
 The derivatives with respect to t of the four branches are
\begin{align}\label{below4}
&-4\beta c^2+2a+\frac{G_3}{G_1}-x'(t)=0,\,\,-4\beta c^2+2a+\frac{G_3}{G_1}-x'(t)=0, \\
&-4\beta c^2+2a+\frac{G_4}{G_2}-x'(t)=0,\,\,-4\beta c^2+2a+\frac{G_4}{G_2}-x'(t)=0, \label{below5}
\end{align}
where $G_3=-8c^3dt-\frac{16c^4dt}{\sqrt{-16c^4dt^2+c^2-d}}$,$G_4=-8c^3dt+\frac{16c^4dt}{\sqrt{-16c^4dt^2+c^2-d}}$.
In order to find a convex point on the contour line,  we set $x'(t)=k$ in Eqs. (\ref{below4},\ref{below5}),  and then get
four equivalent equations of $t$ as follows
\begin{align}\label{below2}
-8c^3dt(2c+\sqrt{-16c^4dt^2+c^2-d})=0,\,\, 8c^3dt(2c+\sqrt{-16c^4dt^2+c^2-d})=0,\\
8c^3dt(2c-\sqrt{-16c^4dt^2+c^2-d})=0,\,\, -8c^3dt(2c-\sqrt{-16c^4dt^2+c^2-d})=0. \label{below3}
\end{align}
Solving  above Eqs. (\ref{below2},\ref{below3}),  we find only one real solution $t=0$, which  means that
 contour line below the asymptotic background is always convex for $0<d<c^2$.

\end{document}